\documentclass[runningheads]{llncs}
\usepackage[T1]{fontenc}
\usepackage{url}
\usepackage[caption=false]{subfig}
\usepackage{graphicx}
\usepackage[strings]{underscore}
\usepackage{enumitem}
\usepackage{hyperref}
\begin{document}
\title{The Transition Matrix - A classification of navigational patterns between LMS course sections}
\titlerunning{The Transition Matrix}
%
\author{Tobias Hildebrandt \and
Lars Mehnen}
%
\authorrunning{T. Hildebrandt and L. Mehnen}
%
\institute{Department of Computer Science, University of Applied Sciences Technikum Wien, Vienna, Austria
\email{\{tobias.hildebrandt, lars.mehnen\}@technikum-wien.at}}
\maketitle              
\begin{abstract}

Learning management systems (LMS) like Moodle are increasingly used to support university teaching. As Moodle courses become more complex, incorporating diverse interactive elements, it is important to understand how students navigate through course sections and whether course designs are meeting student needs. While substantial research exists on student usage of individual LMS elements, there is a lack of research on broader navigational patterns between course sections and how these patterns differ across courses.
This study analyzes navigational data from 747 courses in the Moodle LMS at a technical university of applied sciences, representing (after filtering) around 4,400 students and 1.8 million logged events. Transition matrices and heat map visualizations are used to identify and quantify common navigational patterns.
Findings include that the majority of the analyzed courses exhibit some kind of diagonal pattern, indicating that students typically navigate from the current to the next or previous section. 

\end{abstract}
\section{Introduction and related work}
Learning Management Systems (LMS) like Moodle are increasingly used to support university teaching, and courses created in LMS include more and more interactive elements, such as assignment uploads, automated self-checks with grading and feedback capabilities, voting systems, and forums. As Moodle courses become more complex, incorporating functionalities previously scattered across different systems, instructors want to know whether students are engaging with the courses and its elements as intended and whether course designs can better cater to student needs. Moodle in itself offers, without additional plugins, only rudimentary usage statistics based on static queries and numeric information without incorporating data visualization. 
This research deals with the question, if certain types of navigational patterns can be observed across courses. 

A substantial body of research regarding how students navigate through LMS courses does already exist. However, the majority of this research focus on individual course elements (like quizzes) instead of broader navigational patterns based on course sections. What would be interesting however, is to know how students navigate from section to section, e.g. if they mostly follow the default sequence of sections, or if they often revisit sections, skip sections or jump to specific sections. There is some research that deals with different types of learners, and which navigational behavior they exhibit (e.g.\cite{kuo_behaviour_2021}, \cite{chen_viseq_2020}). However, there is a lack of research dealing with navigational patterns that are course-inherent and not dependent on learner types, and how these patterns differ across courses. 

One reason for this might be, that a majority of research focus on analyzing individual LMS courses, instead of looking at a broader range of courses or performing cross-course analysis. 
Our analysis on the other hand is based on a substantial sample from the Moodle database of our university. The base sample before data preparation includes around 3.4 million events from 5,993 unique students in 2,096 different courses. 
To visualize the navigational behavior we apply heat maps, which have been used in learning analytics before (e.g. \cite{dobashi_heat_2019}). In the area of process mining they have been used for showing transitions between process activities \cite{mahringer_analyzing_2022}.
Researchers have shown how users transition between individual course elements within one course \cite{peach_understanding_2021,shen_understanding_2020,hildebrandt_cross-course_2024}. 
Others have applied heatmaps show transitions between sections of a specific course \cite{kleftodimos_using_2016}.
However, such research has to our best knowledge so far not been conducted to compare patterns between different courses, or for finding cross-course patterns. We are thus interested in the following questions:

\begin{enumerate}
	\item How can the navigation behavior of students moving back and forth between sections of LMS courses be visualized in a heatmap?	
	\item Which different course section navigation patterns exist and how frequent are they? 
	\subitem  Which is the dominant pattern type in cross-course analysis?
\end{enumerate}

\section{Methods}
\subsection{Data retrieval and -engineering}
The data were extracted from the anonymized Moodle database using SQL, and subsequently filtered.
The data processing was conducted using the open-source tool KNIME in conjunction with Python nodes for heatmap generation and quantification (see 
\href{https://hub.knime.com/knme_user/spaces/Public/Transition_Matrix/}{\url{https://hub.knime.com/knme_user/spaces/Public/Transition_Matrix/}}
for the KNIME workflow available on KNIME hub and 
\url{https://github.com/TobiasHildebrandt/Transition-Matrix}
for the Python code).
Table \ref{tab:event_log} shows the structure of the event data.

\begin{table}[htbp]
	\centering
	\caption{Event Log columns}
	\label{tab:event_log}
	\scriptsize
	\begin{tabular}{@{}llllll@{}}
		\hline
		Timestamp & Course Name & CourseID & Section name & Section & UserID \\
		\hline
		2022-08-30 17:25:20.000 & Mathematics for CS & 1 & Section B  & 2 & user1 \\
		2022-09-05 19:26:03.000 & Physics & 2 & Section A & 1 & user2 \\
		2022-09-09 13:27:55.000 & Mathematics for CS & 1 & Section C & 3 & user1 \\
		\hline
	\end{tabular}
\end{table}

The SQL query for the event log was designed to filter the data, exclusively including events associated with student interactions for courses held during the winter semester of 2022. Additionally, it excluded events of courses that had been deleted during the specified time frame and of those with fewer than 100 events, as well as courses for which no grades have been entered. This filtering mechanism aimed to exclude courses potentially created for testing purposes or those not reliant on Moodle for instruction. It further excluded courses that primarily served informational purposes for students and employees. For instance, Moodle is frequently utilized internally for employee training and organizational functions.

\begin{figure*}[htbp]
	\centerline{\includegraphics[width=1\textwidth]{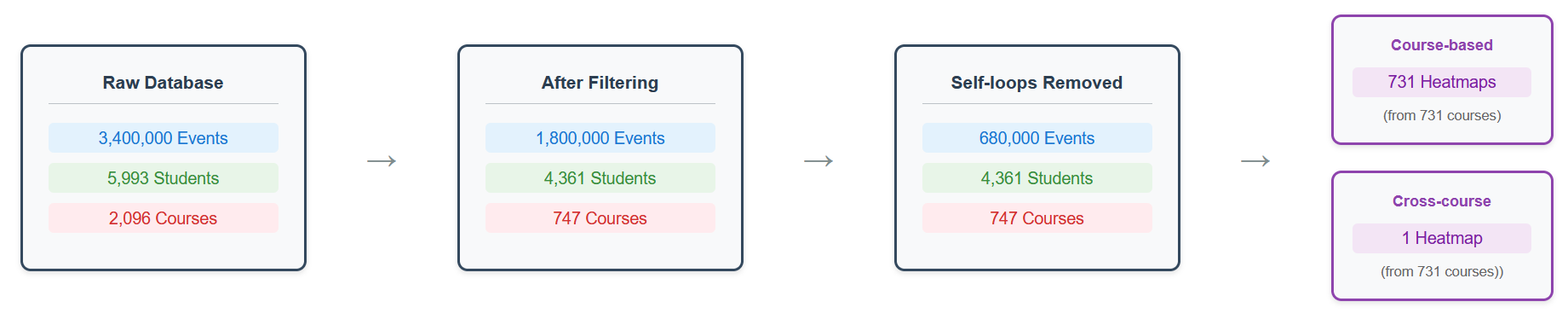}}
	\caption{Data preparation}
	\label{fig:pipeline}
\end{figure*}

After a first pre-filtering of the events, approximately 1.8 million rows remained, involving 4,361 unique students across 747 courses. This translates to an average of approximately 2,400 events per course and around 100 events per student within a course. 
In Moodle, courses can be divided into sections. One section typically corresponds to a specific topic and/or to an in-presence or self study unit.
Self-loops between sections were removed in such a way as that of two subsequent events of the same student in the same course section only the first was kept. Afterward, around 680,000 interactions remained.
The process is schematically presented in Fig \ref{fig:pipeline}.

\subsection{Heatmap creation}
For the heatmap creation, we used Python in conjunction with the packages pandas and seaborn. 
For each individual course a heatmap representing transition frequencies between sections was created as follows:
A transition matrix was created that counts for each source section the total amount of outgoing transitions to each destination section. A transition in this regard is
where for an individual student an event was observed for that student in the source section, followed directly by an event of the same student in one of the other sections (destination sections). It was not important if that second event was observed in the same Moodle session, or if a time period was observed between those two events. 
The rationale behind this was that we were not interested in intra-session transitions, but in navigational behavior over the duration of the whole course. 
The heatmaps can be interpreted as follows:
\begin{enumerate}
	\item X-axis: The Moodle section of origin. This is the section the students have visited previously before changing sections. The sections are ordered from left to right in the same order they appear in the Moodle course.
	\item Y-axis: The destination section. When performing a section change, this is the section the students have visited directly after leaving the previous section. The sections are ordered from top to bottom as they appear in the Moodle course.
\end{enumerate}
The cells are colored using a color transition from white to dark blue. The formula behind the color mapping calculates the number of direct transitions from section X to section Y divided by the total number of transitions going out of section X, multiplied by 100. Thus, each cell has a value between 0 and 100. A white cell indicates that no student has performed the sequence in question (a transition of a visit to section Y after a visit to section X). The darkest shade of blue would indicate a value of 100, meaning that all students that have visited source section X subsequently visit destination section Y (without visiting other sections in between). 731 heatmaps were created, 16 could not be created due to missing section names.

\subsubsection{Cross-course heatmap}
In order to find out if these course-specific patterns also persist when analyzing data in a cross-course-manner, a single heatmap representing all courses has been created. 
Instead of grouping the data to count transitions for individual students between course sections, the data was grouped by course as well as student.
As of course the section names differ between courses, the transition matrix has instead been created based on the sequence numbers of the sections. As not all courses have the same number of sections, the data has been filtered to include only the transitions of the first 10 sections. 

\subsection{Heatmap classification}
\begin{figure*}[htbp]
	\centerline{\includegraphics[width=1\textwidth]{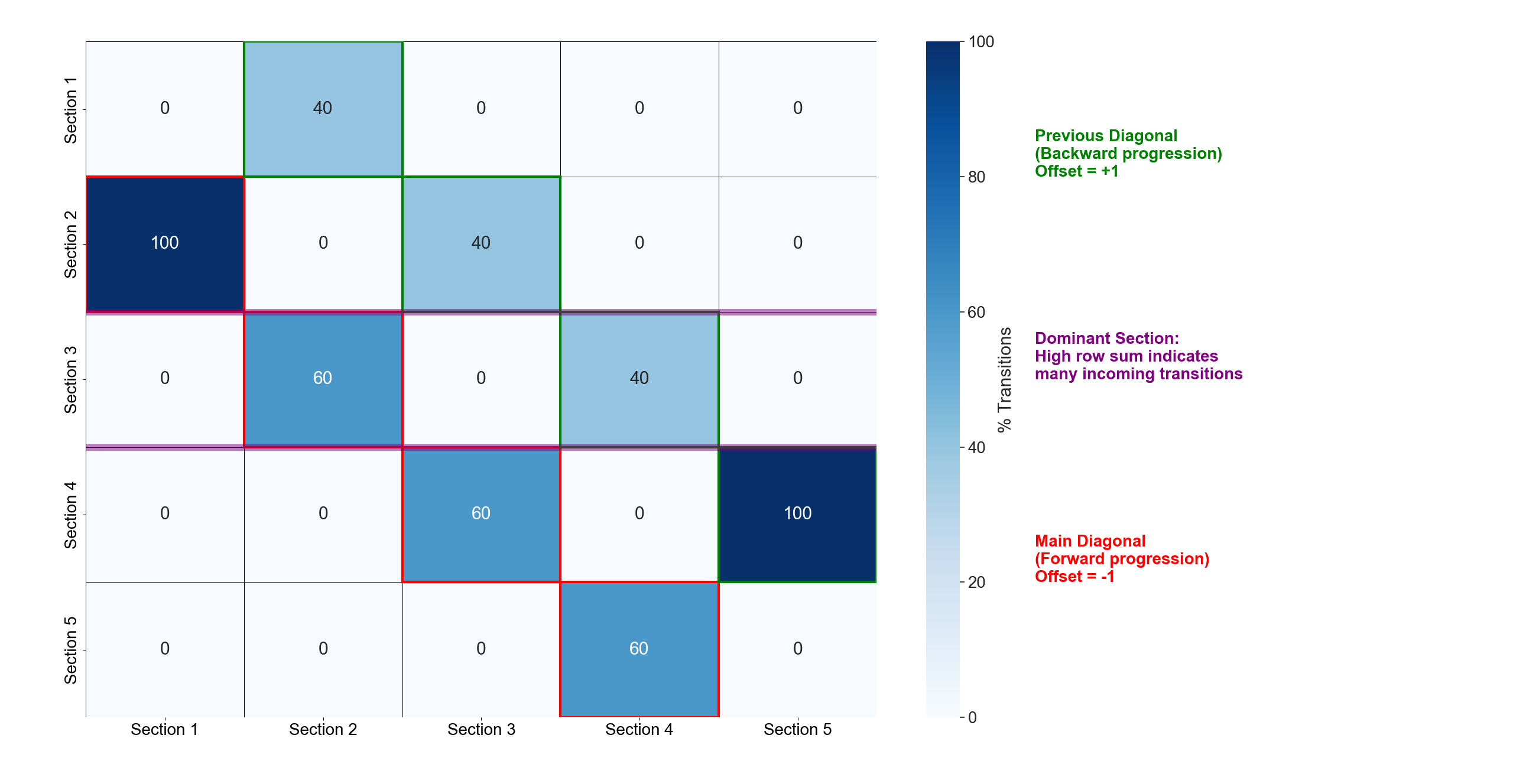}}
	\caption{Illustration of metrics (x-axis: navigation sources, y-axis: navigation targets)}
	\label{fig:metrics_guide}
\end{figure*}

The generated transition matrices where automatically classified according to the navigation patterns they exhibit (see next section for results). The aim was to find out if there are generalizable patterns and how often they occur.
The implementation, available in the aforementioned github-repository as well, works as follows:
The code iterates over all courses, creating the transition matrix in a similar way as during the generation of the heatmaps. For each transition matrix, 6 metrics (illustrated in Fig. \ref{fig:metrics_guide}) are calculated:
\begin{enumerate}
	\item The strength of the main diagonal. This diagonal indicates a forward navigation, in which most students navigate to the next section in the course sequence.  In the matrix, this diagonal is characterized by an offset of -1. The algorithm calculates the percentage of meaningful transitions (defined as over 30 percent of all outgoing transitions) in relation to the total number of elements in the diagonal.
	\item The strength of the previous diagonal (offset 1), indicating a strong pattern of backward movement (revisiting the previous section).
	\item The blended score, a diagonal with an offset of -2 (thus indicating a navigation to the section after the next section). This diagonal, in combination with a main and a previous diagonal, is often found in courses that adhere to the blended learning principle. In such courses, students often jump directly from a self study section to the next self study section (thus skipping the in-presence section in between), or from an in-presence section to the next in-presence section.
	\item The existence of a dominant section (a section that receives most of the incoming traffic throughout the course).
	\item The entropy of the matrix, as a high entropy indicates that no clear 
	pattern is present.
	\item Total main diagonals: the sum of main and previous diagonal.
\end{enumerate}

Based on these metrics, the course matrices are classified into different types as described in the next section.

\section{Results}
\begin{figure*}[htbp]
	\centerline{\includegraphics[width=\textwidth]{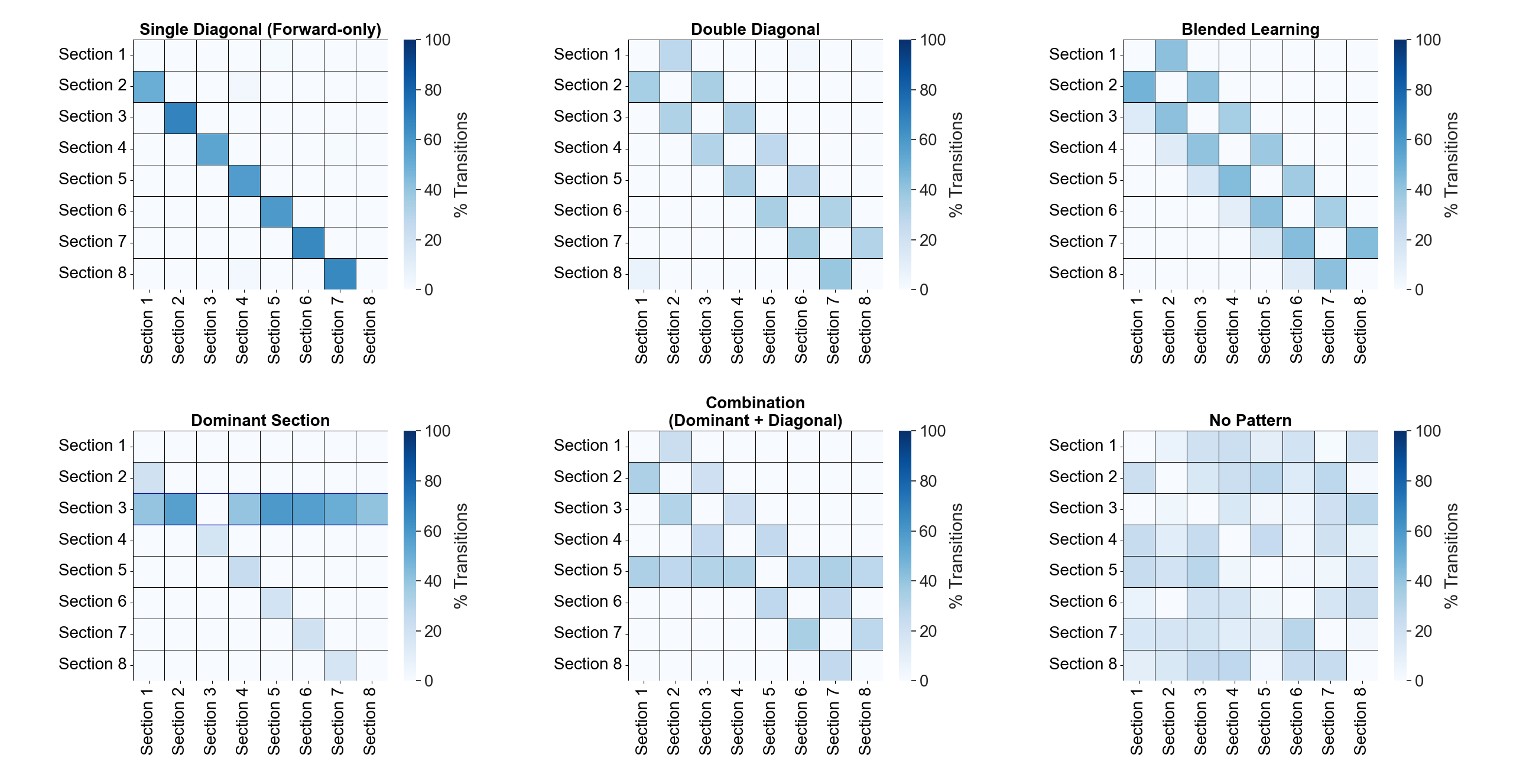}}
	\caption{Types of navigational patterns}
	\label{fig:pattern_guide}
\end{figure*}

When analyzing the 731 heatmaps generated for individual courses, it becomes immediately clear, that most of them follow some kind of diagonal pattern, as discovered by \cite{peach_understanding_2021} for an individual course. However, what can be seen is that such courses with diagonal patterns can be further classified into different sub types. Furthermore it also becomes apparent that there are other courses which do not follow a diagonal pattern, and a few that display no salient trend at all. The found pattern types are presented in Fig. \ref{fig:pattern_guide}.

\subsubsection{Single diagonal}
Courses with such heatmaps are dominated by a single diagonal line, which typically leads from the current section to the next section in the LMS course order, indicating that students follow a progressive sequential path, without the need to often revisit previous section. This pattern is represented by a medium to strong main diagonal combined with a weak previous diagonal, or vice versa.
\subsubsection{Double diagonal}
Double diagonals are another frequently observed pattern. Hereby the students most often go directly from the current section either to the previous or the next section. This indicates courses in which students are encouraged to look up contents from the previous section before continuing. These patterns are a combination of a medium to strong main diagonal and a medium to strong previous diagonal.
\subsubsection{Blended Mode}
This pattern type is classified by a high blended score. There is often a centered “main” diagonal (which again can be a single or a double diagonal), as well as a further diagonal on the left to it. Such structures are often found in courses that adhere to a blended learning paradigm, where students are supposed to go through self-study units before the actual in-presence classes.

\subsubsection{Dominant section}
A further emerging pattern is that of a dominant section (in combination with low to medium total main diagonals), which means that the students often revisit a specific section. This is often the first section of a Moodle course, but not always. This behavior can have different reasons, such as that one Moodle section contains important information that is relevant to the whole course, or alternatively, that all upload fields for assignment submissions are clustered in a specific section. 
\subsubsection{Combination}
A few courses exhibit a combination of high total main diagonals and a dominant section.
\subsubsection{No salient pattern}
A few courses display no strongly discernible pattern. The algorithm assigns this pattern in case of high entropy or if none of the previously described patterns can be attested. 
\subsection{Quantification results}
\begin{figure*}[htb]
	\centerline{\includegraphics[width=0.9\textwidth]{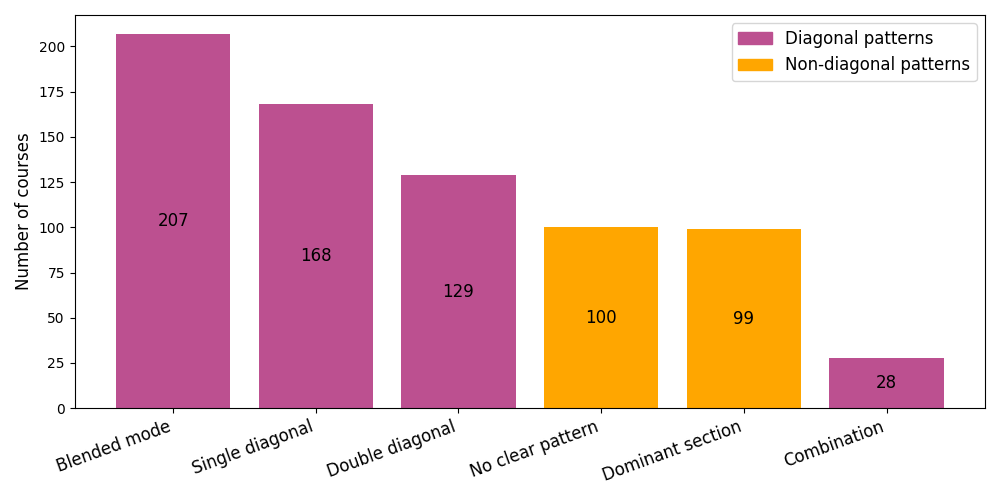}}
	\caption{Quantification of navigational patterns}
	\label{fig:pattern_distribution}
\end{figure*}

After executing the automated pattern classification described previously, the result (see Fig. \ref{fig:pattern_distribution}) shows, that the by far most frequent pattern is that typical for blended courses.
Single and double diagonal are the next frequent types with 23 percent and 17.6 percent respectively. Together with combination (3.8 percent), this means that 72.8 percent of all courses exhibit a clear diagonal pattern of some sort, while 13.5 percent have a dominant section and 13.7 have no clear pattern at all. 

\subsection{Cross-course analysis}
\begin{figure*}[htbp]
	\centerline{\includegraphics[width=0.9\textwidth]{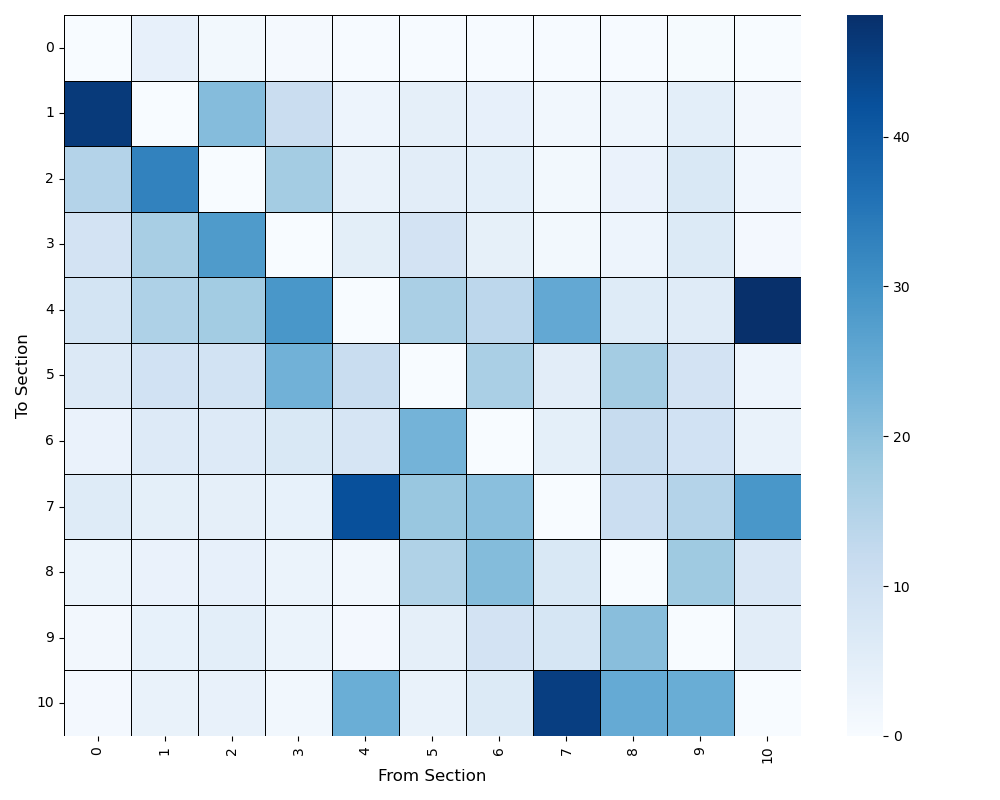}}
	\caption{Cross-course analysis}
	\label{fig:cross}
\end{figure*}

When looking at the cross-course analysis (which creates an average heatmap based on all courses) one can see that the \textit{typical} overall pattern is that of a blended learning course (which is no surprise, given that many of the courses at our institution are designed as such), albeit not a very distinct one. The most protruding pattern is the main diagonal, indicating that students in most cases navigate to the next section in sequence. Further noticeable are a weak blended diagonal, indicating jumps to the section after the upcoming one, and a weak previous diagonal, indicating revisits to the previous section (See Fig. \ref{fig:cross}).

\section{Discussion}
As discussed, the vast majority of courses display some kind of diagonal pattern - in some cases this pattern is more discernible than in others.
When looking at courses without discernible patterns, in many cases this is explainable by the fact that these courses in general do not rely a lot on Moodle (like courses with only few course sections/elements and/or courses that require a lot of personal interaction, e.g. soft skills modules), or by Moodle interaction that is not evenly distributed across sections. 
In general, this classification of course navigation patterns can have implications on the course design. 
If e.g. found out that students often revisit a specific section, it seems that this information is important for the whole course. In such cases it seems sensible to make this information as easily accessible as possible. This can include to make it as easy as possible to navigate through the course and specially to popular sections. In general, diagonal patterns, as found in most courses, may be the desired behavior. Courses with such patterns imply that the information or LMS activity students seek after visiting one section is either in the same, or in adjacent sections. In courses without such clear patterns, when students often navigate to sections that are not adjacent, one possible explanation may be, that the needed information or activities are scattered across the course and potentially should be rearranged. Thus, it would be interesting to especially have a look at the 27.2 percent of courses that do not show a clear diagonal pattern. 
In summary, potential implications for LMS course design could be:
\begin{itemize}
	\item Structure content for a logical flow
	\item Identify hub sections and make them easily accessible (e.g. by \textit{pinning} them)
	\item Avoid unnecessary jumps (by designing for a diagonal navigation pattern)
	\item Review courses without diagonal navigation pattern
\end{itemize}

\subsection{Limitations and Future Work}
The quantification results presented here may not be fully generalizable to other institutional contexts. Our analysis is based on courses from a technical university of applied sciences with a strong emphasis on blended learning pedagogies, which may influence the distribution of navigation patterns observed. However, we argue that the pattern classification methodology itself is transferable across different educational contexts. While the specific percentages of each pattern type may vary in comprehensive universities, liberal arts colleges, or purely online institutions, the fundamental navigation patterns (single diagonal, double diagonal, dominant section, etc.) likely exist across diverse educational settings, even if in different proportions.
While our pattern classification provides empirical evidence for how course structures manifest in practice, we cannot make causal claims about learning outcomes. Our contribution is primarily descriptive—providing a taxonomy and methodology for identifying navigation patterns at scale. Future work could test whether different patterns actually impact learning performance, which would more directly advance learning theory. Further future work could contrast the navigation patterns with the intentions of the respective course designers and educators.

\section{Conclusion}
When designing LMS courses for systems like Moodle, educators want to make sure to design them in a way that best fits the students needs in terms of navigational sequence. Existing research has looked into the sequence in which students use the course elements and sections and more specifically, which section students visit immediately after visiting a specific section. One main result is that one of the main emerging patterns is a double diagonal, where students in many cases navigate to adjacent sections (previous or next sections) - something that others have already investigated for individual courses. However, there is a lack of research investigating such patterns over a wide range of courses, which we have done. We have found further patterns. Specially in courses that follow a blended learning approach, these main diagonals are accompanied by further diagonals from self-study units to in-presence units and vice versa. 
Another pattern that was found was that of dominant section (where students often revisit a specific section), often in conjunction with diagonal patterns. In a few cases, no clear pattern emerged. 
When creating an average heatmap across all courses, the resulting pattern is typical for courses designed for the blended learning principle.

\begin{credits}
\subsubsection{\ackname} The study was financially supported by the Vienna Municipal Department for Economic Affairs, Labour and Statistics as part of the project “Teaching \& Learning Analytics for data-based optimization of teaching and learning processes in courses with blended learning”, which was granted in the 32nd call for Quality Assurance of Teaching at the Viennese Universities of Applied Sciences.
\end{credits}

%
%
%
\bibliographystyle{splncs04}
%
\bibliography{paper2_extended} 
\end{document}